\documentclass{cernrep} 
\usepackage{texnames}
\usepackage[T1]{fontenc}
\usepackage[utf8x]{inputenc}
\usepackage[bookmarks, colorlinks=true, linktoc=page, pdftex, linkcolor=black, citecolor=black, urlcolor=blue]{hyperref}
\usepackage{fancyhdr}
\fancyhfoffset{4 mm}
\fancypagestyle{ARTTITLE}{%
\fancyhf{} 
\lhead{\footnotesize{Proceedings of the 2019 CERN--Accelerator--School course on
\it{ High Gradient Wakefield Accelerators}, Sesimbra, (Portugal)}}
\lfoot{Available online at \url{https://cas.web.cern.ch/previous-schools}}
\rfoot{\thepage\hspace*{3mm}}

   }
\begin{document}
\title{Electron Sources from Plasmas}
 
\author {Brigitte Cros}

\institute{ Universit\'{e} Paris-Saclay, CNRS,  Laboratoire de Physique des Gaz et des Plasmas, 91405, Orsay, France.}

\begin{abstract}
Relativistic electrons are easily generated by self-injection when an intense laser drives a wakefield in a plasma, giving rise to wide electron energy distributions. Several mechanisms  involving additional laser beams or different gas composition or distribution can be used to improve the electron beam quality. These mechanisms are introduced and discussed in the perspective of using laser driven electron sources as injectors for plasma accelerators.

\end{abstract}


\maketitle 
\thispagestyle{ARTTITLE}
\section{Introduction}

A plasma is a reservoir of free electrons moving relatively to ions at the thermal velocity and eventually recombining with them.
An intense laser beam interacting with a plasma can give rise to extremely non-linear situations and expel electrons out of the plasma. The control of involved mechanisms  is crucial for the use of generated electrons.
Requirements for electron beam parameters depend on the specificity of applications, but most of them  require transport and focusing of the beam, together with stable and reproducible parameters from shot-to-shot.
One important case is the achievement of an injector for multi-stage plasma accelerators, where a few-fs  duration electron bunch is required. This has recently been investigated in detail in the frame of the design of EuPRAXIA~\cite{Eupraxia}, a project to build an accelerator based on plasma generating high electron beam quality. 

\begin{figure}[ht] 
  \centering
     \includegraphics[width=14cm]{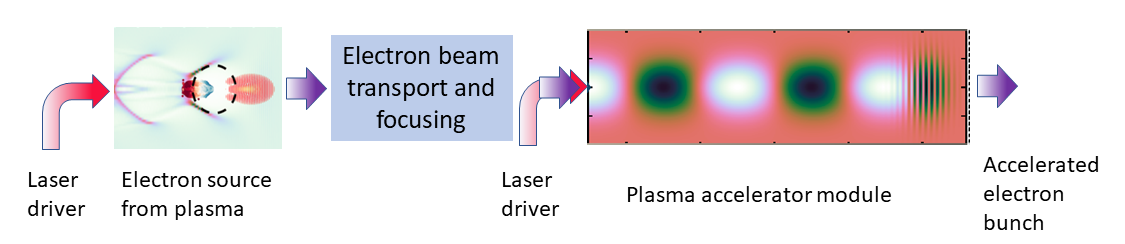}
 \caption{\label{fig:multistage} Schematic design of a multi-stage laser driven accelerator in plasma using electron sources from plasma.
}
  \label{fig:}
\end{figure}

Electron injection into a second plasma is one of the current challenges for the development of plasma based high energy accelerators.  The concept of a plasma based accelerator coupling an electron source to a plasma accelerator stage is illustrated schematically in Fig.~\ref{fig:multistage}.
In this scheme, a laser plasma injector (on the left of  Fig.~\ref{fig:multistage}), generates relativistic electrons bunches as a result of laser driven wakefield in a  non linear regime in the plasma, and does not need  injection from an external electron source.  Electrons from the injector then need to be transported  to and injected into the accelerator operating in the quasi-linear regime for further energy gain without additional electron injection.  The conditions of trapping and  acceleration of electrons in the plasma wave of the injector  have a strong impact on  the~resultant parameters of the electrons at the entrance of the second plasma. Electrons with relativistic energy, 
of the order of 100 MeV or higher are required for transport between stages and  trapping of a~significant number of electrons, typically 10~pC, in the plasma cavity of the accelerator stage. 
 Laser plasma injectors typically operate in a non-linear regime, such that the laser normalized vector potential $a_0  > 1$,  
 corresponding to a peak laser intensity at focus $I_L$ larger than $2.1\times10^{18} \: \mathrm{W/cm^2} $,
  for a laser of wavelength $\lambda = 800$\,nm.

Mechanisms of electron trapping and acceleration in a plasma wave are controlled through laser and plasma properties.
The mechanisms where electrons from the plasma are injected, trapped and accelerated in the plasma wave in the wake of an intense laser pulse 
are designated by the generic term of self-injection. 
When the injection process is simultaneous to acceleration, accelerated electrons with broad energy distributions are generally  achieved at the plasma exit. In order to control or decouple the~processes of electron injection and acceleration, several methods  can be used  such as adding laser beams, tailoring plasma distribution or composition, or a combining the former methods.

 Most of the experimental work to date in the field of laser wakefield acceleration is related to  injector development, and studies the acceleration of small populations of relativistic electrons issued from the bulk of plasma electrons, with which the intense laser interacts.
 These relativistic electron bunches are relatively easy to achieve experimentally with existing laser systems, but their properties remain difficult to control routinely in experimental conditions.
  Over the past fifteen years, successive improvements have been proposed theoretically and demonstrated experimentally. The use of various physical mechanisms provides different levels of control of the electron beam properties at the expense of simplicity.

 External injection and acceleration of electrons in a laser plasma accelerator as described in Fig.~\ref{fig:multistage} has been attempted by only a few groups, and its achievement is complex as it involves femtosecond range timing and micron precision on alignment and stability of beams.  The early work of Amiranoff \textit{et al.} \cite{Amiranoff1998} has demonstrated the need of short electron bunches for injection into a plasma structure, and the extension of the plasma length to the dephasing length in order to fully exploit this scheme. Recent work \cite{Steinke2016} has further demonstrated the potential and challenges of external injection, emphasizing the~need for compact electron beam transport and shaping.

 A selected number of concepts or mechanisms laser wakefield injectors are discussed in the remainder of this report, based on their strong potential for improving the electron beam quality and on  published experimental results.

\section{Electron trajectories in laser driven wakefield}
  
The main features of electron dynamics in the wake of a laser pulse are briefly described in this section. External electrons can be injected and trapped in a plasma wave either in the quasi-linear regime  or in the~non-linear regime. However,  electrons from the plasma are trapped in the plasma wave only in non-linear regimes where there is a mechanism bringing them from rest or thermal velocity to the relativistic phase velocity of the plasma wave.

\begin{figure}[ht] 
  \centering
      \includegraphics[width=15cm]{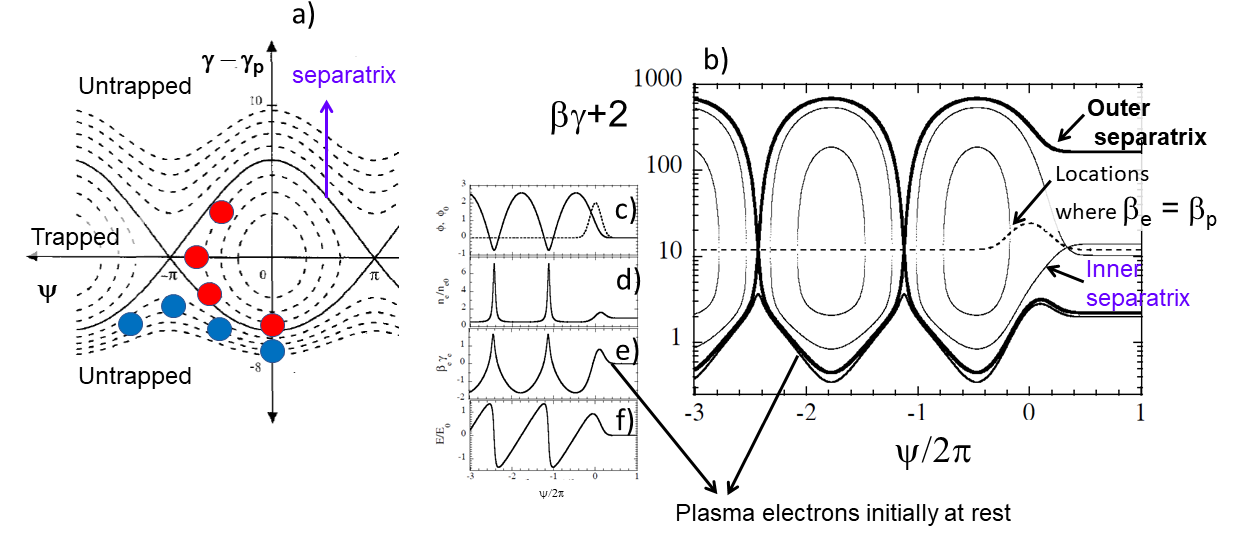}
 \caption{  Electron trajectories in phase space
    (a) in the quasi-linear regime of laser driven wakefield  $a_0 \ll 1$ , wave potential $\phi_0 = 10^{-3}$, relativistic factor of the plasma wave $\gamma_p=20$; blue dots are examples of locations where electrons are running freely, whereas red dots show locations of trapped electrons; and (b) in the non linear regime.  For the non linear regime, graphs in the inset show (c) the plasma wave potential, (d) the density perturbation, (e)~the electron velocity and (f) the accelerating electric field as the function of phase behind the laser pulse which envelope is plotted as a dotted line in (c).
}
 \label{fig:trajectories}
\end{figure}

Electron trajectories in the wakefield can be calculated in one dimension from the equation of motion of an electron along the laser propagation direction. Electron trapping depends on its initial velocity, $ v_e= c \beta_e $, as illustrated in Fig.~\ref{fig:trajectories} for (a) the quasi-linear regime, and (b) the non-linear regime.
Plasma electrons  with  longitudinal velocity component smaller than the phase velocity of the plasma wave, $ \beta_e < \beta_p$ at phase $\psi = 0$,    when the plasma electric field is equal to zero (maximum wave potential), 
are slipping backward  with respect to the plasma wave. Electrons with too low velocity (blues dots in Fig.~\ref{fig:trajectories} (a)) do not gain energy and reach $ \psi = - \pi $ with $\beta_e < \beta_p$, and 
remain untrapped.
Electrons with velocity larger than the phase velocity in the accelerating phase region (red dots in Fig.~\ref{fig:trajectories} (a))
are trapped. $v_p = c \beta_p$ is the phase velocity of the plasma wave.

In the non-linear regime of laser driven wakefield, the density perturbation $n_e/n_0$, ($n_0$ being the~background electron density) behind the laser pulse (Fig. \ref{fig:trajectories} (d)) exhibits spikes separated by plateau along the propagation direction.  The density perturbation can be  expressed as a function of the  electron velocity  as
$n_e/n_0 = \beta_p/(\beta_p-\beta_e)$.
This expression shows that when the electron velocity approaches the~phase velocity the density perturbation approaches infinity and the structure of the plasma wave cannot be coherently sustained, leading to what is called  wavebreaking.  This mechanism provides a way for electrons initially born at rest at the beginning of the interaction to follow trajectories that carries them close enough to the separatrix for trapping to occur.

 \section{Self-injection of plasma electrons}

 The amplitude of the plasma wave and the properties of accelerated electrons in laser driven wakefield are particularly sensitive to the value of the product $\omega_p \tau$,  where $\omega_p$ is the plasma angular frequency associated to the background plasma electron density,  and  $\tau$ is the laser pulse duration. 
For $\omega_p \tau \gg 1$, self-modulated wakefield excitation \cite{Modena1995} results from the modulation of long laser pulses during the growth of plasma waves. For intense enough laser pulses, this regime leads to wavebreaking and continuous electron injection in the plasma cavity, usually over short acceleration  distances due to dephasing in relatively high plasma density. This regime is thus not adequate for the generation of low energy spread electron sources.

\begin{figure}[ht] 
  \centering
     \includegraphics[width=15cm]{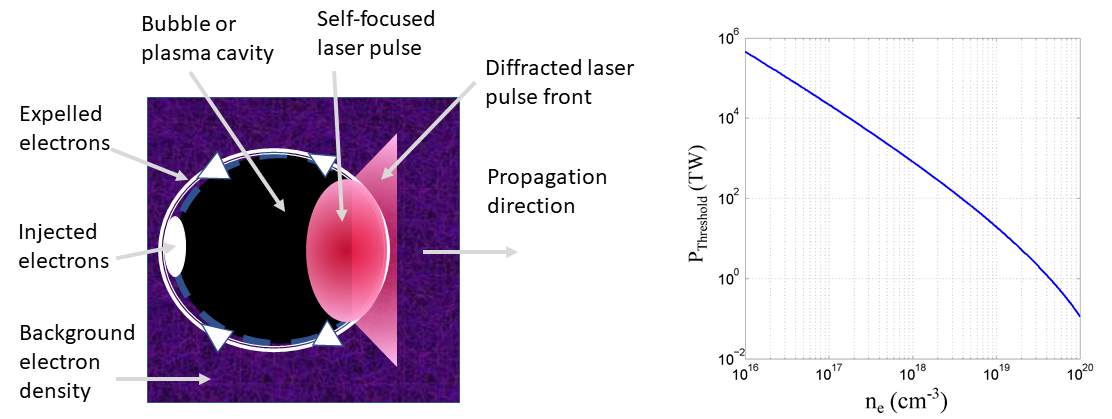}
 \caption{ Sketch of self-injection in the bubble regime showing the main features of laser propagation and electron density distribution in the space containing the direction of laser propagation and a transverse axis; the graph on the~right hand-side shows the value of the power threshold for self-injection as a function of the background plasma density for a laser wavelength of $\lambda=0.8$ $\upmu$m and assuming a Gaussian energy distribution ($\alpha=0.5$).}

  \label{fig:selfinjection}
\end{figure}

Significant improvements can be achieved using short laser pulses, with a pulse duration close to the characteristic response time of the plasma.
Plasma wave generation  close to resonant condition $\omega_p \tau \simeq 1$ provides more efficient plasma wave generation and more controllable electron properties.
For typical laser intensities with $a_0 $ larger than unity, a large amplitude plasma wave is excited in the wake of the laser pulse, creating a plasma cavity or bubble, a volume behind the laser pulse with reduced electron plasma density, as illustrated schematically in Fig.~\ref{fig:selfinjection}.

For sufficiently high laser intensities, typically $a_0>3-4$,  \cite{Mangles2007,Lu2007} electrons at the back of the bubble near the laser axis can be injected in the cavity, where they experience a longitudinal accelerating field. Self-focusing  of  the intense part of the laser pulse leads to  lengthening of the bubble. As a consequence, some electrons passing through the bubble during its lengthening gain enough energy to be injected in the plasma wave \cite{Saevert2015}.

In experiments, for a given laser system, a density threshold is commonly observed, below which no electron beam is generated. The highest beam quality (in terms of low energy spread, and stability) and electron energy are achieved for a working point just above the density injection threshold \cite{Mangles2012}.  However, the plasma density needs to be high enough to accelerate a significant charge and to reach electron self-injection conditions sufficiently early in the interaction for acceleration to occur after injection. 
Mangles \textit{et al.} \cite{Mangles2012} determined a threshold on laser power to achieve self-injection, taking into account the fraction of laser energy, $\alpha$, contained in the full-width at half-maximum (FWHM) of the laser pulse  as a function of plasma electron density. The variation of this power threshold  is plotted as a function of plasma density in Fig.~\ref{fig:selfinjection}, indicating the range of laser and plasma parameters where self-injection takes place.

Non-linear lengthening of the first plasma period behind the laser pulse was measured by S\"avert \textit{et al} \cite{Saevert2015} in gas jets, providing experimental evidence of  self-injection  caused by the lengthening of the~first plasma period.
Electron self-injection and acceleration to energies of hundreds of MeV have been  achieved experimentally by several groups \cite{Mangles2004,Geddes2004,Faure2004,Banerjee2012,Saevert2015,Froula2009}. The generation of electron beams with a charge in the range 50-100 pC,  spread over a large energy range is typically  observed.
The highest electron energy reached after self-injection was  combined with laser guiding in a plasma channel, achieving 6 pC of charge at 4.2 GeV in a single stage with an energy spread as low as 6\% (RMS) and a divergence below 1 mrad \cite{Leemans2014}.

Self-injection is the most straight-forward injection scheme and the simplest to  implement in laser driven wakefield acceleration  \cite{Esarey2009}. However, in the self-injection scheme, electron injection and trapping in the plasma wave continues until  either the amplitude of the wakefield is reduced through beam loading, that is the electric field generated by the trapped electrons reduces and finally cancels the wakefield, or laser energy depletion lowers the laser amplitude below the thresholds for self-focusing and self-injection. Self-injection is thus usually characterized  by a large energy spread of the accelerated electron bunch. 

Within the self-injection scheme highest quality electron beams are achieved in the bubble regime \cite{Pukhov2002} where the laser with relativistic laser intensity is focused to a sphere of radius shorter than the plasma wave wavelength. The laser and plasma parameter range for the generation of high-quality electron beams is  narrow \cite{Saevert2015} and shot-to-shot fluctuations are usually large.
Electron trapping and acceleration by an expanding bubble was
analysed theoretically \cite{Kalmykov2009}:
 electrons that would normally slip back can be trapped as the bubble radius changes.
   In this case bubble expansion was caused by defocusing: defocusing followed by focusing is proposed as a way to stop injection.

\section{Triggered injection of plasma electrons}
     
     Self-injection is easily achieved but continuous injection of a  large number of electrons makes the process noisy and good beam quality is difficult to achieve at selected energy range. Some control over the~injection of plasma electrons can be achieved by triggering the injection process, by either using one or more additional laser beams, or by acting on the plasma density profile or composition to start injection at a specific phase value during the propagation of the main laser pulse.
     
     \subsection{Colliding-pulse injection} 
The colliding-pulse injection scheme is based on the use of two laser pulses propagating toward each other and crossing in the plasma: a drive laser pulse with an intensity below the self-injection threshold creates a plasma wave and a second laser pulse, called the injection pulse, is used to locally increase the~intensity and trigger electron injection into the plasma wave \cite{Umstadter1996}. The interaction of the two laser pulses creates a beat-wave pattern, which can, if the intensity is high enough, give a kick to some electrons and put them on a  closed orbit. 
This technique was used to produce electron bunches in the  energy range of 100 to 200~MeV, with a relative energy spread as low as
 $\sim1 $\% and a beam charge of the order of 
 $\sim10$ pC \cite{Rechatin2009,Geddes2016}.

Colliding-pulse injection allows control of the location of injection by adjusting the location where laser pulses interfere and consequently provides a mean to control the acceleration length and the electron energy \cite{Cormier2014}. 
The injection volume is controlled by tuning the intensities of the laser pulses, and its decrease provides the smallest energy spread and charge. This technique imposes particular challenges on the laser system. It requires the use of two laser pulses with good stability for both laser pulses to achieve low fluctuations of electron bunches properties. Furthermore, in a directly counter propagating geometry, laser systems would require extra protection to prevent propagation of laser energy backward in the laser amplifier chain.

\subsection{Density gradient injection }

\begin{figure}[ht] 
  \centering
   \includegraphics[width=14cm]{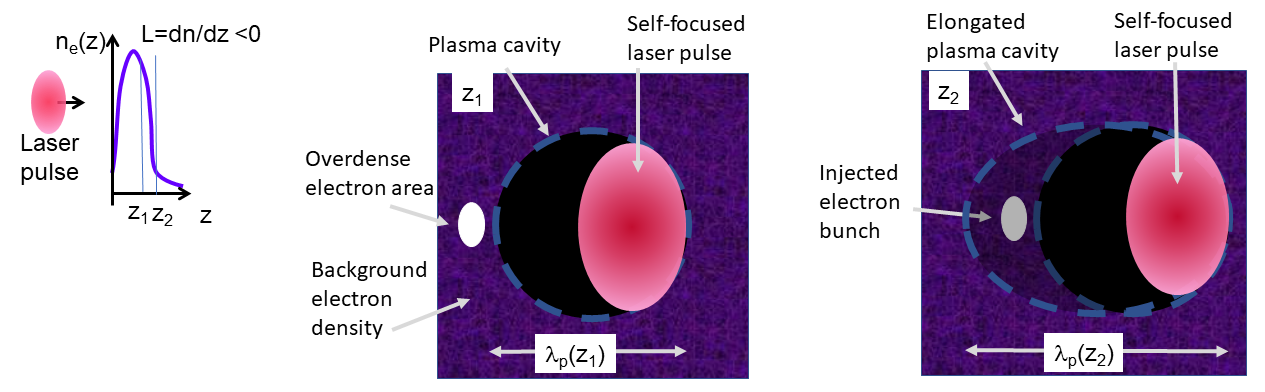}
 \caption{ Cartoon of gradient injection in a negative density gradient as shown on the left hand side of the figure. As the laser propagates from $z_1$ to $z_2$, the background plasma density decreases; electron distributions are schematically shown at $z_1$ and $z_2$.
}
  \label{fig:grad_inj}
\end{figure}

The dependence of the wake phase velocity on the plasma density allows to control injection though the plasma density profile. In particular, a negative plasma density gradient will lower the wake phase velocity and trigger the injection \cite{Bulanov1998}, as illustrated in Fig.~\ref{fig:grad_inj}.
\\

 The  plasma wave wavelength, $\lambda_p (z) \sim 33/ \sqrt{n_e(z)}$, 
so that a decreasing density gradient leads to an increase of $\lambda_p$.
Along the propagation direction $z$,  the wake phase velocity, $ v_p(z,t) = \omega / k  = - \delta _ t \psi / \delta_z \psi $, can be evaluated using the variation of the phase along the density gradient, $ \psi(z) = k_p(z) (z - v_g t) $, were $v_g$ is the group velocity of the plasma wave and $ k_p = 2\pi / \lambda _p$.
The~phase velocity can be written as 
\begin{displaymath}
v_p(z,t) = \frac{v_g}{1+ (z-v_gt)\delta_z k_p/k_p} \, .
\end{displaymath}
Behind the laser pulse, $z - v_g t <0$, and in the decreasing density gradient, $L = \delta_z k_p/k_p <0$, so that the phase velocity decreases, i.e. 
 the wake slows down as the laser propagates down the gradient. This lowers the trapping threshold and some background electrons can then be injected in the wakefield.

Injection through plasma density profile tailoring has been investigated both numerically and experimentally for a number of plasma density profiles and/or laser parameters. 
Depending on the density gradient length $L$,  two regimes can be distinguished. For $L > \lambda_p$, the back of the  plasma wave phase velocity slows down, lowers the threshold for wavebreaking and causes trapping of background electrons at a specific position. For a sharp density transition ($L \leq \lambda_p$), a sudden increase in the plasma wavelength is produced leading to rephasing of the plasma electrons into the accelerating phase of the plasma wave.

A sudden change in plasma density triggers and stops self-injection. To achieve this in practice
an object is placed in the gas flow of a gas jet to create a shock associated to a change of gas density over a short distance.
As laser ionisation occurs over a short time scale,  the plasma profile follows  gas profile \cite{Geddes2008,Schmid2010,Buck2013, Burza2013}.
The density gradient can also be produced by using a secondary laser pulse \cite{Chien2005,Faure2010}, 
or by using two separate overlapping gas jets \cite{Hansson2015,Wang2016}. In those cases, the location of the density gradient can be changed relative to the density profile and the energy of the electrons can thus be tuned.
Density gradient injection has been  used to achieve electron beams with energy in the range $25-600$~MeV, with relative energy spreads of the order of $1-25$\%, and with a charge in the range $1-1000$~pC \cite{Geddes2008,Faure2010,Buck2013,Wang2016}.

Compared to self-injection in a homogeneous plasma,  higher quality, nearly dark-current free electron beams are obtained by  injection in a steep density down-ramp. Injection probabilities close to 100\% and improved (with respect to self-injection) shot-to-shot stability are generally achieved in experiments. Shot-to-shot fluctuations as low as $\sim 2$\% of the electron energy, $\sim 6$\% of the charge and pointing stability better than 1 mrad have been reported \cite{Gonsalves2011}. 
Experimental studies indicate that a good control of the plasma density profile is needed, small changes can affect the electron beam quality and the stability. 
One of the major difficulties of this injection technique is the realization and the control of the sharp density down-ramp.

\subsection{Ionisation-induced injection}

\begin{figure}[h!] 
  \centering
     \includegraphics[width=8cm]{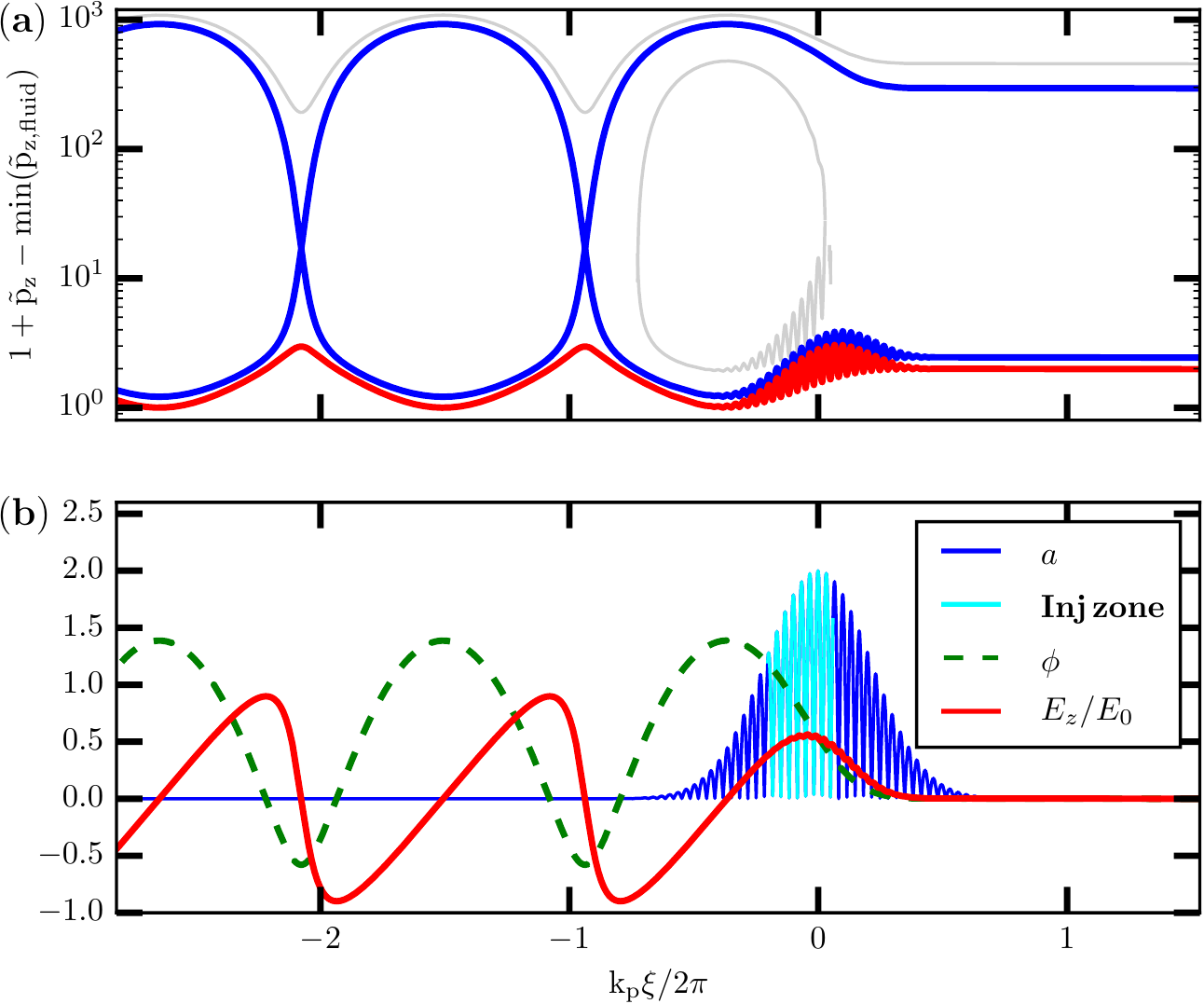}
 \caption{ a) Electron trajectories in phase space  showing background untrapped electron trajectory (in red), the~separatrix 
(in blue) and a typical trajectory for electrons ionized and trapped  in  the first period of 
the wakefield (in gray);  b) Laser pulse with normalized vector potential $a_0 = 2$, accelerating field, $E_z$, and  wake potential $\phi$ . An~electron ionized in the region 
colored in cyan will be injected and trapped in the wakefield; from~\cite{Lee2017}.
}
  \label{fig:iiitrap}
\end{figure}

Adding a small amount of gas with high atomic number atoms, such as nitrogen, to the bulk of low atomic number atoms, such as hydrogen, provides another way to control electron injection into the~plasma wave. 
Outer shells of the high-atomic number atoms are ionized together with the low-atomic number atoms, at relatively low intensities (typically below $I_L = 10^{16}$ $\mathrm{W/cm^2}$) so that free electrons appear  at the front of the laser pulse and contribute to the plasma wake in a way similar to electrons coming from the main gas component. Inner-shell electrons are ionized closer to the laser intensity peak and are more likely to be injected in the wake. A large gap between ionising potential for outer and inner shell electrons favors the generation of electrons at different positions as electron ionisation from different levels follows the evolution of the laser envelope: an electron born close to the peak of the laser envelope (cyan colored zone  in Fig.~\ref{fig:iiitrap}) will be trapped.
This electron experiences a larger difference of potential than an electron born at the front of the laser pulse 
and is thus more likely to be injected into the wake. Chen \textit{et al.} \cite{Chen2012} describe how an electron born stationary inside the wake is turned around by the potential of the non-linear wake provided that 
$a_0 \gtrsim 1.7(1-n_e/n_{c})$, where $n_c$ is the critical density associated to the laser frequency.

Ionisation induced injection was first observed in beam driven  plasma wakefield experiments where  narrow energy spread electron beams  from a plasma were achieved in the blow-out regime \cite{Oz2007}.
The first observation of ionisation induced injection in a laser driven wakefield experiment occurred in hydrogen filled capillaries: 150 MeV electrons with $<10\%$ energy spread were measured driven by low laser intensity $a_0<1$ \cite{Rowlands-Rees2008}. Simulations performed to analyse data demonstrated that ionisation induced injection of impurity atoms from the surface of the capillary walls, associated to laser pulse intensity growth during guided propagation such that the trapping threshold could be reached, could explain the appearance of peaked electron spectra \cite{Kamperidis2014}. Subsequent experiments were able to utilise this technique with  dopant species  introduced on purpose to produce electron bunches with charge in the~range
$1- 100$~pC and relative energy spreads of the order of
5  to 10~\% \cite{McGuffey2010,Pollock2011,Golovin2015,Hansson2016,Hafz2016}.

Ionisation induced injection provides a way to trigger electron injection into the accelerating phase. However  injection continues as long as the laser intensity is high enough to ionise inner shell electrons, leading to accelerated electrons with broad energy distributions \cite{Pak2010,McGuffey2010,Desforges2014}. 
Main advantages  of  ionisation induced injection technique are that it is  easy to implement experimentally, the charge can be increased compared to self-injection\cite{Desforges2014}, it can operate at a lower intensity threshold compared to self-injection \cite{Chen2012} (which is beneficial for a potential increase in repetition rate) and can potentially reduce the transverse emittance \cite{Chen2012}.

\section{Combination of several mechanisms }

Ionisation induced injection scheme can be used in a long plasma  to scale up  to higher energies and in that case produces broad electron energy distributions. However, combined with either density gradient injection or colliding pulse injection, it is a way to increase the locally injected charge. 
This combination of mechanisms offers tunability of charge, energy, energy spread, and emittance by increasing the number of control parameters, while increasing complexity of implementation and  constraints on stability.

The energy spread can be reduced by tuning the laser intensity to be above the threshold for ionisation injection only in a small volume of interaction. Using carefully controlled  laser intensities, the~maximum intensity achieved through pulse compression can be large enough to achieve electron trapping for only a small distance of laser propagation or duration \cite{Kamperidis2014}. It is also possible to use structured targets, composed of a region containing a gas mixture for the injection followed by a region of pure gas to further accelerate injected electrons \cite{Pollock2011,Golovin2015}, or tailored density profile to combine ionisation-induced injection to a density gradient \cite{Hansson2016}.

 An example of a case where ionisation induced ionisation combined to density tailoring is used to improve beam properties is illustrated here, using PIC simulation results obtained  with the code WARP~\cite{vay2012}. 
\begin{figure}[ht] 
  \centering
   \includegraphics[width=12cm]{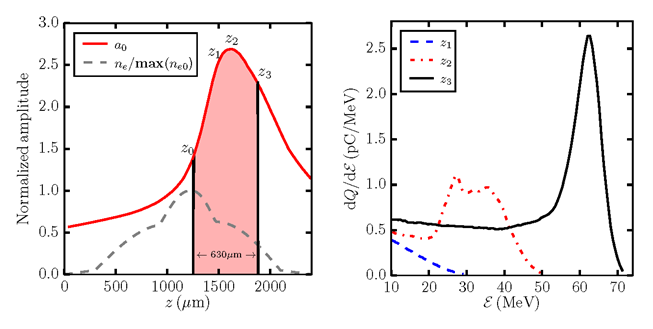}
 \caption{ Left  hand graph:  Evolution of the maximum laser amplitude on axis, $a_0$ as a function of position $z$ along the laser propagation axis; the dashed line shows the longitudinal electron density profile inside a gas cell; the~shaded area  of length ∼630~$\mu$m indicates injection range;  labels are defined as follows:  $z_0$, injection zone; $ z_1$,  position where injection begins; $z_2$,  position where $a_0$ is maximum; $z_3$, the position where injection stops. Right hand graph: energy spectra of electrons  accelerated at the 3 positions,$z_1$, $z_2$, and $z_3$. 
}
  \label{fig:iiigradient}
\end{figure}
Simulations of ionisation induced injection were  performed with the density profile shown by the dashed line in Fig.~\ref{fig:iiigradient}, with a maximum electron density on axis equal to $ 7.8 \times 10^{18}$~cm$^{−3}$. This profile is typical of profiles achieved in a gas cell by  modelling of gas flow ~\cite{audet2018}. In this example, the~plasma with a total length  of 2.4~mm, was ionised by the laser propagating in a gas composed of  99\%~H2 (hydrogen) and 1
\%~N2(nitrogen).  The laser is described by a Gaussian function with a normalized vector potential  of maximum value in vacuum at focus, $ a_0= 1.1$.
The evolution of $a_0$ (red plain curve in Fig.~\ref{fig:iiigradient}) shows that its maximum value  is larger in plasma than in vacuum due to self-focusing in the varying density profile. In this example electron injection starts at position $z_0$ where $a_0 $  has reached the value 1.5,  and stops at $z_3$, when the plasma density is too low for electron trapping, resulting in a~total injection length of $630~\mu$m.

Corresponding spectra are plotted at the three positions along the propagation on the right hand side of Fig.\ref{fig:iiigradient}. The shape of the electron energy distribution is evolving as the electron bunch propagates, leading to a reduced energy spread. 
At $z_1$, just after the start of the injection process, the energy is low and the charge density decreases continuously with energy. At $z_2$, a broad peak with central energy around 32~MeV is formed, evolving into a distribution with a larger amplitude peak above 60~MeV at $z_3$.

This is the result of a redistribution of electrons during the acceleration process in presence of beam loading: electrons injected earlier contribute to the bulk of the peak in z3; they experience smaller accelerating fields compared to later injected electrons due to beam loading, so that later injected electrons can catch up during the propagation with the initially injected ones and  contribute to populate the~peak.

With the chosen laser plasma parameters, simulation results produce an electron bunch with a~peak energy of 65.7 MeV, a FWHM relative energy spread  of 13.1\% and a charge of 43.6 pC.
 The moderate power laser pulse restricts the
injection to only ionization induced injection and a focal position in the~descending gradient of the longitudinal density profile allows a slow growth of the vector potential,
a0, delaying the ionization processes, and resulting in the shortening of the injection range as compared to the~plasma length. In this parameter range, beam loading effects are responsible for two distinct phenomena: the inhibition of the injection process and the homogenization of the energy distribution of the trapped electron bunch.
A detailed analysis of electron dynamics in the injection and acceleration processes can be found in~\cite{Lee2018}.

Combination of ionisation induced injection and profile tailoring at large laser power has been explored by a few groups. Couperus et al.~\cite{Couperus2017} have shown that focusing a 200~TW laser power at the~back of  a gas jet composed of
99\% helium and 1\% nitrogen, peak plasma density $3.1\times 10^{18}$~cm$^{-3}$, electrons bunches with 0.25~nC charge, peaked around 250~MeV with a relative energy spread FWHM 15\%.  A  large amount of electrons is injected at large laser power, and the measured accelerated charge was show to grow with the square root of laser power.

Tailored density profile with a long, lower density plasma tail have been tested experimentally: the~injection zone is thus followed by  and acceleration zone, where plasma waves are both driven by the~same laser pulse, with or without assistance from ionisation.
First tests of ionisation injection followed by an accelerator tail ~\cite{Pollock2011} were performed focusing a 40~TW laser pulse inside a  double gas cell. A 3~mm long injector doped with 0.5\% nitrogen, was followed by a 5~mm long helium filled accelerator, resulting in an electron bunch close to 500~MeV, with charge 1.4~pC per MeV, a relative energy spread FWHM of 5\% and 2.3 mrad divergence. The lower density in the accelerator section prevents self-injection and favors longer dephasing length, resulting in larger output energy.

\section{Summary}

Achieved parameters  at plasma exit, for electron sources shorter than 10~mm, driven by  20~TW  to 500~TW laser power,   are summarized in Table~\ref{tab:dempar}.
\begin{table}[h]
\caption{Range of  parameters for electron plasma sources}
\label{tab:dempar}
\centering\small
\begin{tabular}{lll@{}}
\hline\hline
\bfseries Beam parameters  & 
 \bfseries Typical Range \\\hline
Energy  & 50-500 MeV \\
Charge  & 0.1-100 pC \\
Bunch length &  3-20 fs \\
Repetition rate  & < 1 Hz \\
Energy spread &  1-15\%  \\
Transverse normalized emittance &1-5 mm mrad \\
Transverse size & 3-10 µm \\
Transverse divergence & 1-10 mrad \\
                                      \hline\hline
\end{tabular}
\end{table}
It should be noted that  these parameters are usually not achieved  all together and different experimental settings are required to optimise specific parameters. 
 High charge (100 pC), combined with low energy spread (5\%rms) and low emittance (1mm mrad), seems to be within reach of identified mechanisms but still remains to be demonstrated experimentally.

Several concepts and techniques are being studied  to optimise electron parameters according to the requirements for transport and applications; fast progress has been achieved over the last 10 years.  A~combination of several schemes, such as gradient and ionisation induced injection, looks like the most promising path to explore in order to achieve the whole set of parameters in a reliable way.
Controlled injection into a plasma wave accelerator will open the way to accelerator development. Preliminary results \cite{Steinke2016} 
 were achieved  in a plasma channel, driven in the quasi-linear regime, using an electron source driven by laser in a gas jet.
The acceleration of externally injected electrons  in plasmas will greatly benefit from the optimisation of reliable electron sources from plasmas. 

\pagebreak

\end{document}